\documentclass[twocolumn,showpacs,preprintnumbers,amsmath,amssymb]{revtex4}

\usepackage{graphicx}
\usepackage{dcolumn}
\usepackage{bm}
\usepackage{grffile}

\usepackage{amssymb,amsmath}

\begin{document}
\title{Site specific spin dynamics in BaFe$_{2}$As$_{2}$: tuning the ground state by orbital differentiation}

\author{P. F. S. Rosa,$^{1,2}$ C. Adriano$^{1}$, T. M. Garitezi$^{1}$, T. Grant$^{2}$, Z. Fisk$^{2}$, R. R. Urbano$^{1}$, and P. G. Pagliuso$^{1}$}

\affiliation{$^{1}$Instituto de F\'isica \lq\lq Gleb Wataghin\rq\rq,
UNICAMP, Campinas-SP, 13083-859, Brazil.\\
$^{2}$University of California, Irvine, California 92697-4574,
USA.}

\date{\today}

\date{\today}

\begin{abstract}
 The role of orbital differentiation on the emergence of superconductivity in the Fe-based superconductors remains an open question to the scientific community. In this investigation, we employ a suitable microscopic spin probe technique, namely Electron Spin Resonance (ESR), to investigate this issue on selected chemically substituted BaFe$_{2}$As$_{2}$ single crystals. As the spin-density wave (SDW) phase is suppressed, we observe a clear increase of  the Fe 3$d$ bands anisotropy along with their localization at the FeAs plane. Such an increase of the planar orbital content interestingly occurs independently on the chemical substitution responsible for suppressing the SDW phase.  As a consequence, the magnetic fluctuations combined with the resultant particular symmetry of the Fe 3$d$ bands are propitious ingredients to the emergence of superconductivity in this class of materials.
\end{abstract}

\pacs{74.70.Xa, 76.30.-v}
\maketitle

Structural parameters in low-symmetry layered systems,
such as iron pnictides/chalcogenides, cuprates and heavy fermion (HF) compounds
have played an important role in determining both symmetry 
and dimensionality of the magnetic fluctuations \cite{Stewart, Mont, Piers}. 
An immediate consequence of a structural control (or tuning) parameter is its influence on the crystalline electrical field (CEF) effects, which are often relevant in determining the ground state in strongly correlated materials. Particularly for HF superconductors, the $f$-electrons strongly hybridize with conduction electron ($ce$) bands and the single-ion anisotropy, mostly defined by the CEF effects, influences the magnetic fluctuations at the Fermi surface (FS) \cite{HF1,HF2, HF3, HF4, HF5}.
In fact, the superconducting transition temperature ($T_{c}$) trendly scales with the lattice parameter ratio c/a for some HF systems \cite{HF1,HF5}. Yet for the high-$T_c$ cuprates, several structural parameters, such as the Cu-apical oxygen distance and the bond length between Cu and the in-plane oxygen, have been suggested as control parameters of $T_{c}$.
Remarkably, the apical oxygen distance to the CuO$_{2}$ plane also display a reasonable scaling with $T_{c}$ \cite{cuprates1, cuprates3}.  In the same fashion, the Fe-pnictide distance has been recently suggested as a similar control parameter tuning the CEF levels in the Fe-based compounds and, consequently, changing the orbital contributions to the bands at the Fermi level \cite{Granado,KH}.
Moreover, it is notorious that this orbital interplay is indeed crucial to the formation of the spin-density wave (SDW) phase due to the $\mathbb{Z}^{2}$ symmetry break (or $xz/yz$ orbitals). Nonetheless, the ultimate role of orbital degrees of freedom on the emergence of supercondutivity in Fe-pnictides materials remains an unsolved puzzle.

In this regard, several spin probes such as electron spin resonance (ESR), nuclear magnetic resonance (NMR), muon spin rotation ($\mu$SR), M\"ossbauer spectroscopy and magnetic neutron scattering, can in principle be used to explore the CEF and orbital differentiation effects on the spin dynamics of a system and their possible influence over its superconducting order parameter. 
The observation of homogeneous resonances by such techniques in strongly correlated metals allows one to obtain the imaginary part of the spin susceptibility $\chi^{''}(\textbf{q}, \omega)$. 
In particular, the ESR absorption line of a paramagnetic local probe is a measure of $\chi^{''}(\textbf{q}, \omega)$, which in turn is directly related to the spin-lattice relaxation rate (1/$T_{1}$) of the localized resonating spins through the $ce$ and thus to the lattice \cite{Abragam, Poole}.
In a general approach for single-band metals, the thermal broadening $b$ of the ESR linewidth $\Delta H \simeq 1/T_{1}$ is linear in temperature and is defined as:
\begin{equation}
b \equiv \frac{d\left(\Delta H\right)}{dT} = \frac{\pi k_B}{g \mu_B} \langle J^2_{fs}({\textbf{q}}
)\rangle\eta^2\left(E_F\right)~\frac {K(\alpha)}{(1 - \alpha)^2}~,  \label{1}
\end{equation}

\noindent which is the so-called Korringa relaxation \cite{ref19}. Here, $\langle J_{fs}(\textbf{q})^{2}\rangle^{1/2}$ is the effective exchange interaction between the local moment and the $ce$ in the presence of $ce$ momentum
transfer averaged over the whole FS \cite{ref21}, $\eta(E_F)$ is the ``$bare$" density of states (DOS)
for one spin direction at the Fermi level, $k_B$ is the Boltzmann constant, $\mu_B$ is the Bohr magneton, $g$ is the local moment $g$-value and $K(\alpha)$ is the Korringa exchange enhancement factor due to electron-electron exchange interaction \cite{ref28,ref29,note}. The $g$-shift $\Delta g$ can be written as \cite{ref18}:

\begin{equation}
\Delta g \equiv g - g_{insulator} = J_{fs}({\textbf{0}}) ~\frac {\eta\left( E_F \right)}{1 - \alpha}~.
\label{2}
\end{equation}

At the local moment site, the $g$-shift probes the $ce$ polarization ($\textbf{q} = 0$). On the other hand, the Korringa rate probes the $ce$ momentum transfer ($0\leq \textbf{q} \leq 2 k_F$) averaged over the FS \cite{ref21}. Thus, from Eqs. \ref{1} and \ref{2} one obtains information about the density of states at the Fermi level as well as the strength and \textbf{q}-dependence of the exchange interaction.

Previous studies on the series Ba$_{1-x}$Eu$_{x}$Fe$_{2}$As$_{2}$ has shown that 
$\langle J_{fs}^{2}(\textbf{q}) \rangle^{1/2}$ decreases with decreasing Eu content \cite{Rosa}. The scaling of $\langle J_{fs}^{2}(\textbf{q}) \rangle^{1/2}$ with $T_{SDW}$ suggests that the electron bands with appreciable overlap with the Eu$^{2+}$ 4$f$
states become more anisotropic (less $s$-like) as the SDW phase is suppressed. Moreover, these electron bands also move, in average, further away from the Eu$^{2+}$ sites in real space \cite{Rosa}. Therefore, in order to generalize such a scenario and further investigate the role of symmetry changes in highly occupied 3$d$ bands in superconducting members of this family, we perform in the present report a systematic ESR study on Ba$_{1-x}$Eu$_{x}$Fe$_{2-y}M_{y}$As$_{2}$ single crystals (with $M$ standing for transition metals Co, Cu, Mn, Ni, and Ru). It is worth pointing out that the Ba and FeAs planes are probed by the ESR active Eu$^{2+}$ and Mn$^{2+}$/Cu$^{2+}$ paramagnetic spins, respectively. The ($x = 0.2$; $y = 0$) compound was taken as reference for the evolution of the spin dynamics as a function of $M$ substitution ($y$ content) because it is the lowest Eu concentration presenting Korringa relaxation. 

\begin{figure}[!ht]
\includegraphics[width=0.5\textwidth,bb=0 0 750 1075]{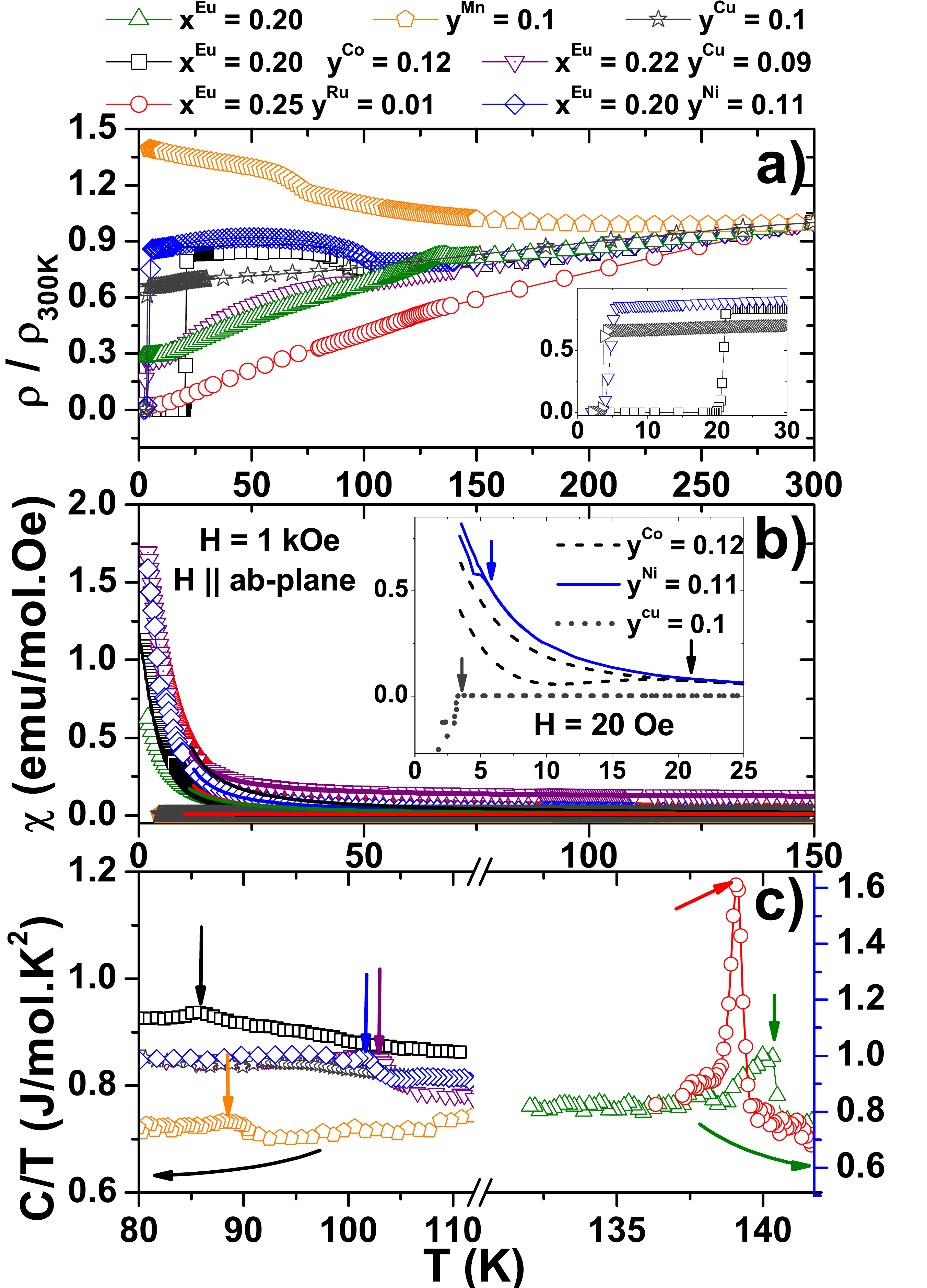}
\caption{Thermodynamic properties of the Ba$_{1-x}$Eu$_{x}$Fe$_{2-y}M_{y}$As$_{2}$ single crystals. The insets show a) the zero electrical resistivity at the superconducting transition and, b) zero-field cooling/field cooling magnetic susceptibilities.}
\label{Fig1}
\end{figure}

Starting off with the macroscopic physical properties, Fig. \ref{Fig1}a
displays the temperature dependence of the normalized electrical resistivity for the studied single crystals.
Room-$T$ values of $\rho$($T$) vary within the range $0.2-0.8$ m$\Omega$.cm. Except for $M$ = Ru, which displays the smallest substitution content ($y=0.01$), the electrical resistivity increases at $T_{SDW}$, as typically found for doped BaFe$_{2}$As$_{2}$ samples\cite{Review}. Besides, for the chosen reference compound ($x = 0.2$; $y = 0$), a metallic behaviour is observed down to $T_{SDW}$ where a sudden drop at $137$ K is identified in its dataset. A suppression of the SDW phase with lower transition temperatures of $135$ K, $102$ K, $98$ K, and $86$ K is clearly observed as Fe is replaced by transition metals $M = $Ru, Ni, Cu, and Co, respectively. Superconducting state also emerges for Co and Ni substitutions with $T_{c}=22$ K and $6$ K, respectively (inset to Fig. 1a). 
In good agreement, the SDW phase is also suppressed for BaFe$_{2-y}M_{y}$As$_{2}$ (i.e., $x=0$) in comparison with the parent BaFe$_{2}$As$_{2}$ with T$_{SDW}\simeq 140$ K.This can be noted for the $y_{Mn} = 0.1$ compound with $T_{SDW}$ = $78$ K. In particular, the SDW phase is fully suppressed for $y_{Cu}=0.1$ giving way to superconductivity at $T_{c}=3.8$ K (inset to Fig. 1a).
                                                                                                                                             
Fig. \ref{Fig1}b shows the in plane magnetic susceptibility as a function of temperature for $H=1$ kOe. For all samples, $\chi(T)$ is well fit to the Curie-Weiss law plus a $T$-independent Pauli term, $\chi(T) = \chi_{0} + C/(T-\theta_{CW})$ (solid lines). We obtain an effective moment $\mu_{eff}\approx 8\mu_{B}$ from the Eu$^{2+}$ ions and an electronic spin susceptibility of $\chi_{0} = 2(1) \times10^{-3}$ emu/mol-Oe for all compounds. The superconducting transition at low-field ($H = 20$ Oe) is defined by the arrows in the inset of Fig. 1b. It is worth noting that, for $M$ = Co and Ni, the diamagnetic response below $T_{c}$ is overcome by the Eu$^{2+}$ paramagnetic one. Fig. \ref{Fig1}c shows the $T$-dependence of the specific heat for the selected single crystals.
For the reference compound with $x=0.2$ and $y=0$, the SDW transition is defined by a jump at $137$ K (green arrow). In agreement with the $\rho$ data of Figs. 1a-1b, we observed a suppression of the SDW phase as Fe is substituted by transition metals $M$.

Now we turn our attention to the ESR results on the selected single crystals in the paramagnetic (PM) state ($T>T_{\rm{SDW}}$). A single asymmetric Dysonian ESR resonance is observed for all compounds, meaning that the skin depth is
smaller than the sample dimensions \cite{Dyson}. Nonetheless, as presented in Fig. 2, the X-Band ($\nu \simeq 9.5$ GHz) ESR spectra at room-$T$ become Lorentzian-like after gently crushing the single crystals in order to improve the signal to noise ratio. From fits to the resonances using the appropriate admixture of absorption and dispersion (solid lines), we obtained the $T$-dependence of both the linewidth $\Delta H$ and $g$-values presented in Fig. 3.

A linear (Korringa) increase of $\Delta H$ with increasing-$T$ is observed for the Eu$^{2+}$ ESR signal in the PM state. From linear fits to $\Delta H (T)$ above $T_{\rm{SDW}}$ we extracted the values of the Korringa rate $b \equiv \Delta H/\Delta T$. It is evident that $b$ decreases systematically for any of the transition metal substitution. On the other hand, when $x=0$ and $M$= Mn$^{2+}$ or Cu$^{2+}$, the ESR signal from these magnetic probes broadens as the temperature is lowered indicating that a competition between the spin-spin and spin-$ce$ interactions must be occurring. In the high-$T$ range, the Korringa-like relaxation dominates though. Thus, linear fits to $\Delta H (T)$ shown in the inset of Fig. 3a allowed us to obtain $b_{Mn^{2+}} = 1.1(2)$ Oe/K and $b_{Cu^{2+}}=3.5(2)$ Oe/K, respectively.

\begin{figure}[!ht]
\includegraphics[width=0.5\textwidth,bb=0 0 800 575]{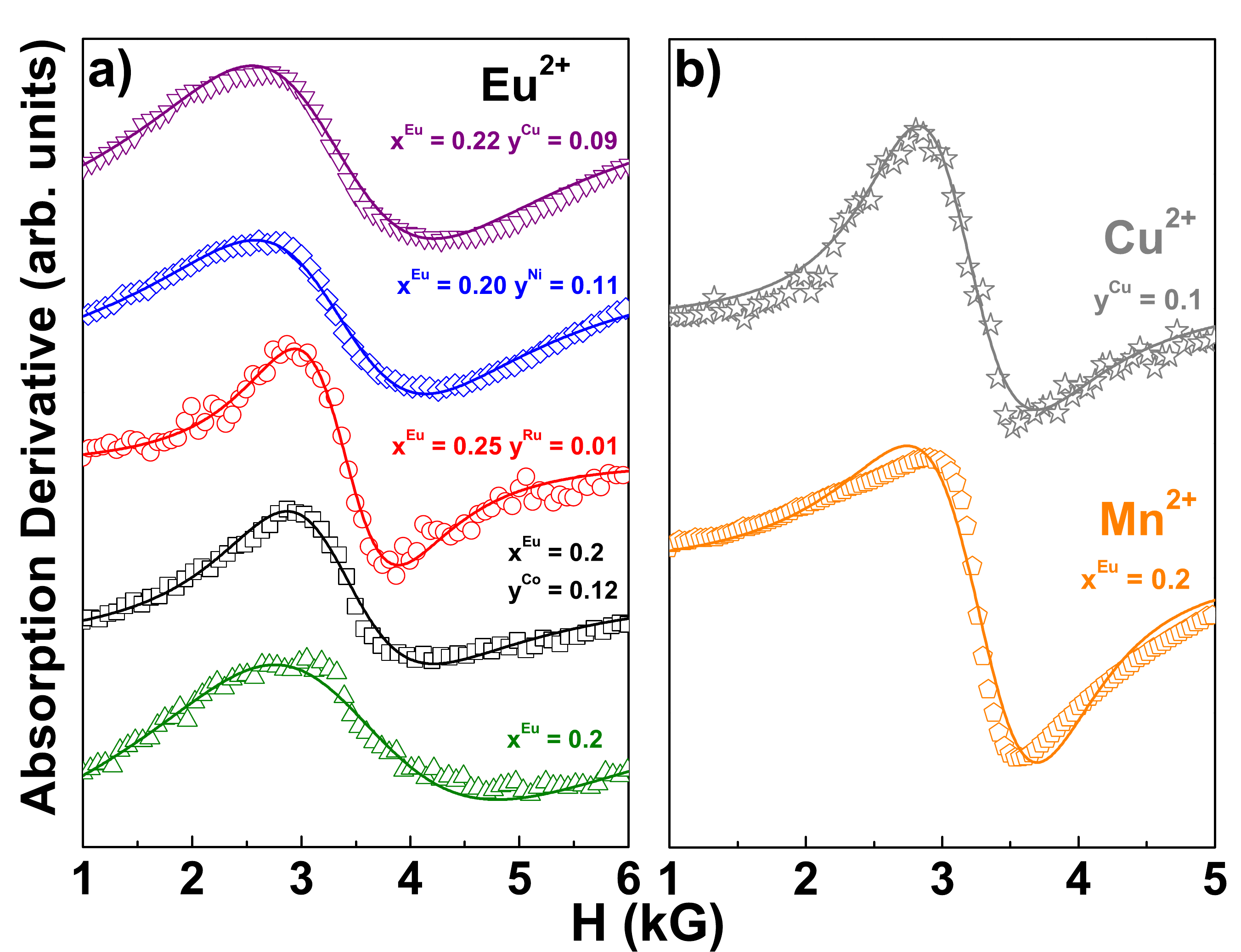}
\caption{X-Band ($\sim$ 9.5 GHz) ESR spectra of Ba$_{1-x}$Eu$_{x}$Fe$_{2-y}M_{y}$As$_{2}$ single crystals at room temperature.}
\end{figure}

The $T$-dependence of the Eu$^{2+}$ ESR $g$-value in the normal state is presented in Fig. 3b. 
Before the long range SDW order sets in, a roughly $T$-independent $g\simeq$ 2  is commonly observed for the Eu$^{2+}$ ESR signal as previously suggested by reports on EuFe$_{2}$As$_{2}$, Eu$_{0.5}$K$_{0.5}$Fe$_{2}$As$_{2}$ and EuFe$_{2-x}$Co$_{x}$As$_{2}$ \cite{Dengler, Pascher, Ying}.
In fact, the high-$T$ Eu$^{2+}$ $g$-values of our Ba$_{1-x}$Eu$_{x}$Fe$_{2-y}M_{y}$As$_{2}$ single crystals are, in average, equal to 2.04(4) and independent of Eu$^{2+}$ content $x$. The highest precision in $g$-value achieved for those samples with the narrowest ESR linewidth $\Delta H$ will not be far from $g$ = 2.05(3).  As it can be verified in Fig. 3b, such a value is indeed a reasonable mean $g$-value for all $x$ for $T > T_{\rm{SDW}}$. 

Now, based on the framework discussed in Ref. \cite{Rosa}, we are confortable to neglect significant changes in the DOS as well as any evolution of the electron-electron interaction as a dominant contribution to the Korringa suppression observed as a function of $M$ substitution.
First of all, an electronic spin susceptibility $\chi_{0} = 2(1) \times10^{-3}$ emu/mol-Oe about one order of magnitude larger than the estimated magnetic Pauli susceptibility from the heat capacity coefficient $\gamma$ is observed for all compounds.
In the presence of such an enhancement, the host metal $ce$ spin susceptibility can be approximated by
\begin{equation}
\chi_0 = 2 ~\mu^2_B~\frac{\eta\left(E_F\right)}{1 - \alpha},
\label{3}
\end{equation} 
where $\alpha$ accounts for the e-e interaction, (1 - $\alpha$)$^{-1}$ is the Stoner enhancement factor and $\eta\left(E_F\right)$ is the "bare" DOS for one spin direction at E$_{F}$ \cite{ref23,ref24}.
Now, assuming that the enhancement in $\chi_{0}$ is only due to e-e interaction, an $\alpha \approx$ 0.85(5) can be estimated. Consequently, based on Ref.\cite{ref29} one can determine the corresponding K($\alpha$) = 0.2(1).
Then, using the relation $(\pi k_B/g \mu_B) = 2.34 \times10^4$ Oe/K and replacing $\Delta g
 \simeq$ 0.05(3),  $\eta (E_F) = 3.34$ states/eV mol-spin and $b$ values from Table 1 into Eqs. \ref{1} and \ref{2}, we are were able to extract $J_{fs}({0})= 2(1)$ meV and $\langle J_{fs}^2(\textbf{q}) \rangle^{1/2}$ for all compounds. These values are also presented in Table 1.

\begin{figure}[!ht]
\includegraphics[width=0.55\textwidth,bb=0 0 900 1100]{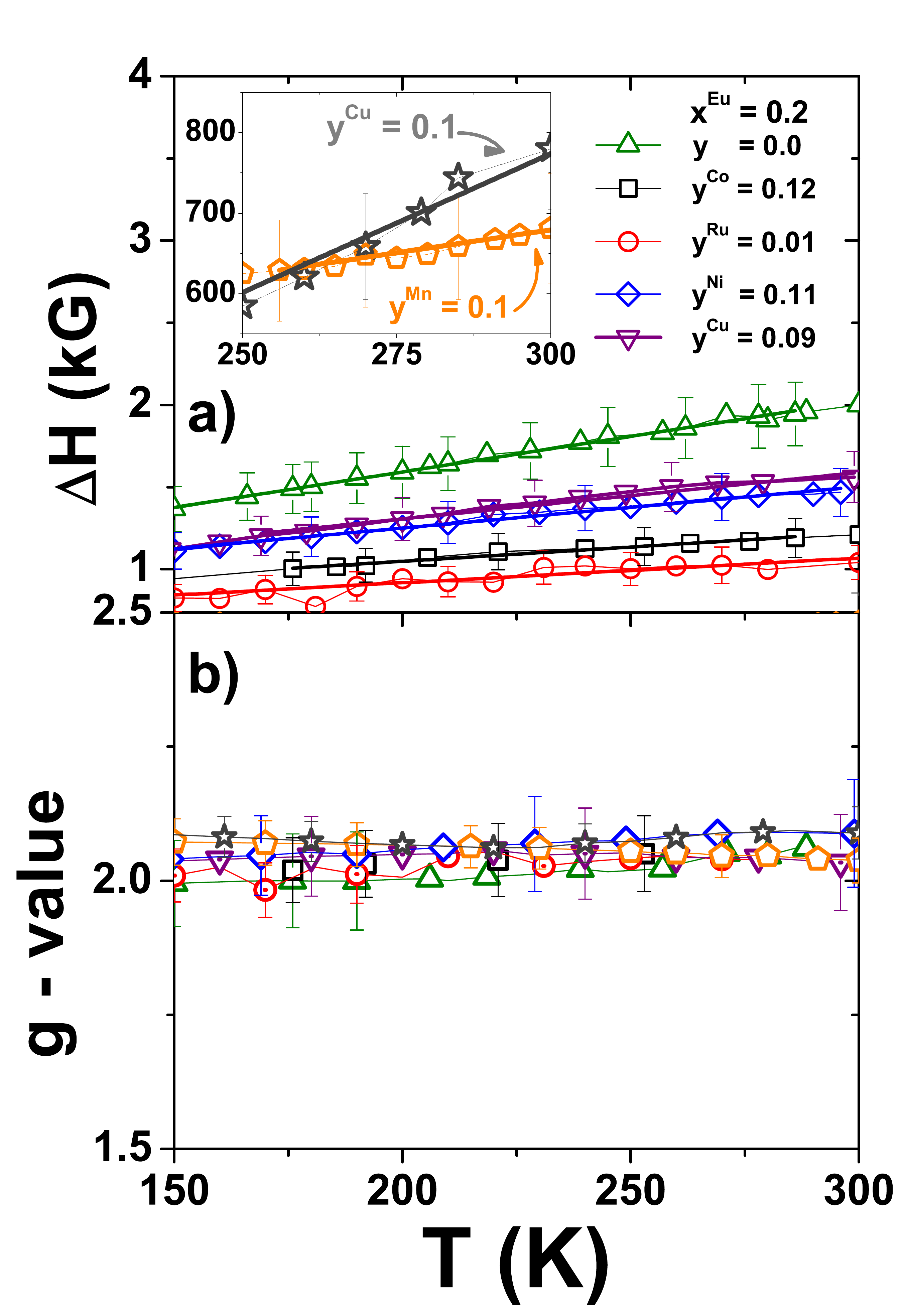}
\caption{$T$-dependence of a) the ESR linewidth $\Delta H$ and b) the $g$-values measured for the powderized samples.}
\label{evolution}
\end{figure}

\begin{table*}
\label{Tab_J} \caption{Experimental ESR Korringa rate $b$, the g-values, the $g$-shift $\Delta g$ \cite{shift}, the exchange interaction $J_{fs}({0})$ and calculated $\langle J_{fs}^2(\textbf{q}) \rangle^{1/2}$ for all samples studied in this work.}
\begin{centering}
\begin{tabular}{c c c c c c}
\hline
\hline  
Sample      &  $b$ (Oe/K) & $g$-value & $\Delta g$  & $J_{fs}({0})$ meV & $\langle J_{fs}^2(\textbf{q}) \rangle^{1/2}$ (meV)  \tabularnewline
\hline 
Ba$_{0.8}$Eu$_{0.2}$Fe$_{2}$As$_{2}$ (ref. compound)    &  4.3(2) & 2.04(4) & 0.05(3)  & 2(1) &     1.4(8) \tabularnewline
BaFe$_{1.9}$Cu$_{0.1}$As$_{2}$                                       &  3.5(2) & 2.08(4)  & 0.08(3) & 2(1) &    1.2(6)   \tabularnewline
Ba$_{0.78}$Eu$_{0.22}$Fe$_{1.91}$Cu$_{0.09}$As$_{2}$  & 2.8(2) & 2.05(4)  & 0.06(4) & 2(1) &     1.1(8)  \tabularnewline
Ba$_{0.8}$Eu$_{0.2}$Fe$_{1.89}$Ni$_{0.11}$As$_{2}$       & 2.5(2)  & 2.04(4) & 0.05(3) & 2(1) &      1.0(8)  \tabularnewline
Ba$_{0.8}$Eu$_{0.2}$Fe$_{1.88}$Co$_{0.12}$As$_{2}$      & 1.7(2) & 2.05(3)  & 0.06(3) & 2(1) &      0.9(7)  \tabularnewline
Ba$_{0.75}$Eu$_{0.25}$Fe$_{1.99}$Ru$_{0.01}$As$_{2}$ & 1.5(2) & 2.04(3)  & 0.05(3) & 2(1) &      0.8(7)  \tabularnewline
BaFe$_{1.88}$Mn$_{0.12}$As$_{2}$                                   &  1.1(2) & 2.04(3)  & 0.05(3) & 2(1) &     0.7(6)  \tabularnewline
\hline
\hline  
\end{tabular}
\par\end{centering}
\end{table*}

Based on the data of Figs. 1-3, the diagrams shown in Fig. 4 summarize the behaviour of $T_{\rm{SDW}}$ and Korringa relaxation rate $b$ against $M$ substitution.
When the ESR spin probe is out of the Fe-As plane (Fig. 4a), we observe that both $T_{\rm{SDW}}$ and $b$ values are suppressed with respect to their values for Ba$_{0.8}$Eu$_{0.2}$Fe$_{2}$As$_{2}$ as Fe is substituted by $M$. Nevertheless, when the spin probe is in-plane (Fig. 4b), an opposite behaviour takes place: the Korringa rate $b$ increases as the SDW phase is suppressed.
Therefore, these results combined evidentiate that the relative value of $\langle J_{fs}^{2}(\textbf{q}) \rangle^{1/2}$ is clearly diminishing with $y$ when the probe is out of the Fe-As plane in Ba$_{0.8}$Eu$_{0.2}$Fe$_{2-y}M_{y}$As$_{2}$. On the other hand, $\langle J_{fs}^{2}(\textbf{q}) \rangle^{1/2}$ increases as $T_{\rm{SDW}}$ decreases when the probe is in turn in the Fe-As plane. Given that $J_{fs}({q})$ is the Fourier transform of the spatially varying exchange interaction, by suppressing the SDW phase, the electron bands with appreciable overlap with the Eu$^{2+}$ 4$f$ states become more anisotropic (less $s$-like). Moreover, these bands, in average, move further away from the Eu$^{2+}$ sites in real space, i.e., they assume a more planar/$xy$-orbital character. 
On the other hand, it is also clear that there is no unambiguous systematics between the magnitude of the SDW phase suppression and that of the Korringa rate, specially for the Ru substitution. This is because the particularities in the distortion of the eletrocnic bands may behave differently for each transition metal. For instance, if the chemical substitution is not coherent, interference can occur among the $3d$ bands and consequently the suppression of the Korringa rate $b$ might be more effective than that of $T_{\rm{SDW}}$ \cite{PJ}.

\begin{figure}[!ht]
\includegraphics[width=0.5\textwidth,bb=0 0 800 1050]{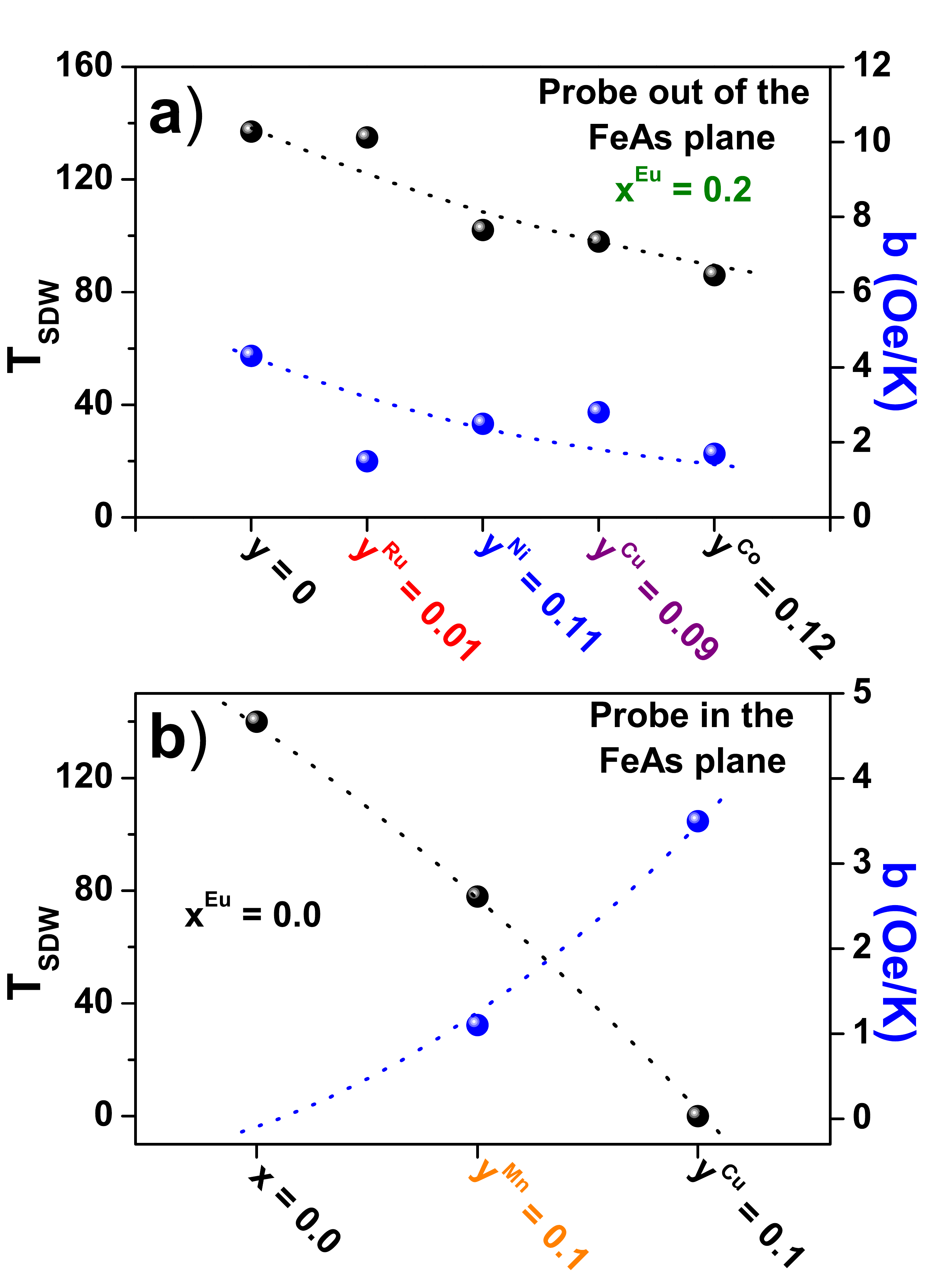}
\caption{Spin-density wave transition temperature $T_{\rm{SDW}}$ and Korringa rate $b$ as a function of transition metal substitution $M$ when the ESR spin probe is a) out of the Fe-As plane and, b) in the Fe-As plane.}
\label{diagram}
\end{figure}

It is imperative noticing that our findings are in complete agreement with the data on the concentrated Eu regime. A slower Korringa rate $b$ has also been observed when K or Co were introduced into EuFe$_{2}$As$_{2}$ \cite{Pascher,Ying,Fernando}. Furthermore, DFT+DMFT calculations have recently shown that orbital differentiation is crucial when magnetic correlations become relevant in BaFe$_{2}$As$_{2}$ \cite{KH}.
By decreasing the iron-pnictogen distances, the $xy$ orbital occupancy increases, which in turn weakens the Fe magnetic moment. This affects the Fermi surface since a decrease in the occupancy of the $xy$ orbital results in the increase of the hole pocket size, as seen by ARPES studies in several $A$Fe$_{2}$As$_{2}$ families \cite{Setti, ReviewARPES}.

In order to give this reasoning a more graphical view, we have explored how the exchange interaction $J$ between the 3$d$ conduction electrons and the local spin ESR probes may be affected by the orbital differentiation of the 3$d$ bands. As such, we assume a scenario where $J$ is proportional to the overlap between the atomic orbitals. Although simplified, this procedure can qualitatively capture the main features of the orbital differentiation through minute changes in $J$.
Therefore, when the paramagnetic probe is out of the Fe-As plane, one can calculate the squared overlap between Eu 4$f$ and Fe 3$d$ orbital wavefunctions, $\langle\Psi_{4f}|\Psi_{3d}\rangle^{2}$, as a function of the distance between the Eu and Fe ions, namely $z_{0}$ \cite{model}. Ultimately, as shown in Eq. \ref{1}, the crucial result within this scenario is that the Korringa relaxation rate $b$ is directly proportional to $\langle\Psi_{4f}|\Psi_{3d}\rangle^{2}$.

\begin{figure}[!ht]
\includegraphics[width=0.5\textwidth,bb=0 0 800 600]{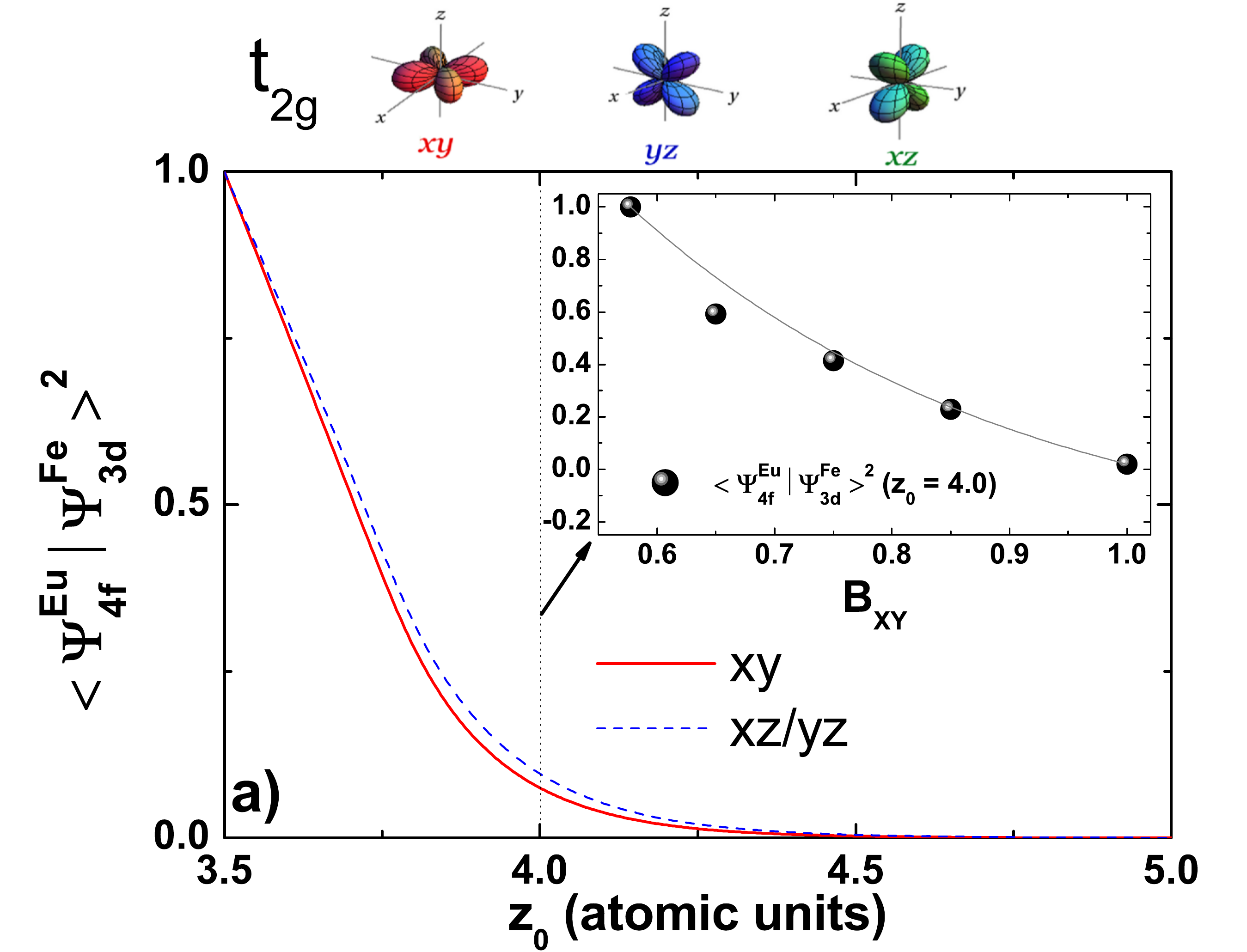}\\
\includegraphics[width=0.5\textwidth,bb=20 0 800 600]{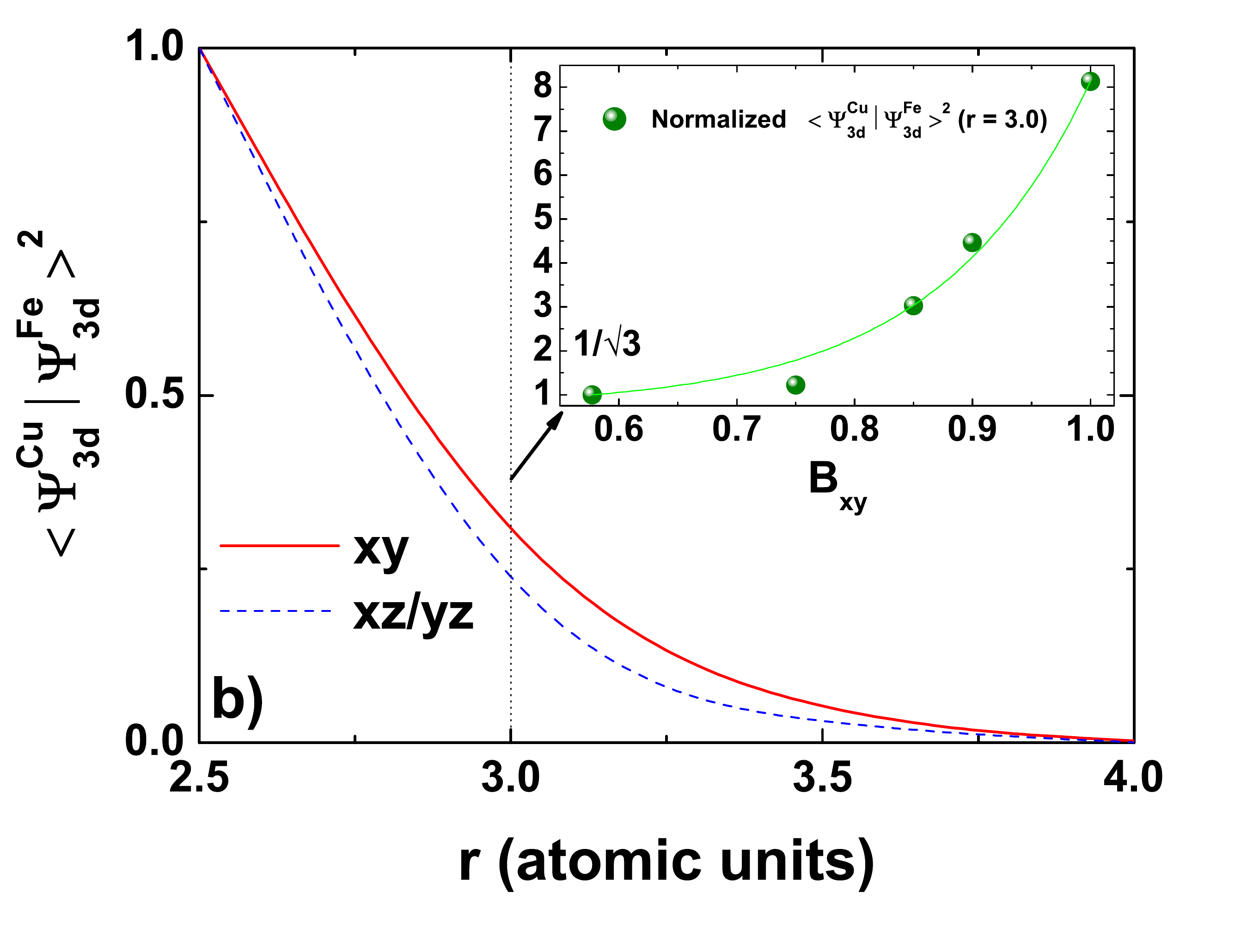}\\
\vskip -0.5cm
\caption{Squared overlap between the Eu 4$f$ isotropic wavefunction and the $t_{2g}$ Fe 3$d$ wavefunctions $xy$ (B$_{XY} = 1.0$) and $xz/yz$ (B$_{XY} = 0.0$) when a) the probe is at the Fe-As plane;  b) the probe is out of the Fe-As plane.}
\label{Theory}
\end{figure}

Let us first consider the overlap between an isotropic Eu 4$f$ configuration and the Fe t$_{2g}$ orbitals: $xy$ and $xz/yz$. The triplet orbitals were chosen here due to their predominant role on the FS \cite{KH}. The squared overlap $\langle \Psi_{4f}^{Eu}|\Psi_{3d}^{Fe}\rangle^{2}$ between the Eu 4$f$ and the Fe 3$d$ wavefunctions as a function of $z_{0}$, normalized by the atomic radius of Eu$^{2+}$ ($z_{0}=3.5$ atomic units), is presented in Fig. 5a.
We note that, when the planar $xy$ orbital component is the dominant one (solid red line), its squared overlap with the 4$f$ wavefunction decreases slightly faster than that with only $xz/yz$ orbital components (dotted blue lines). One would expect such behaviour simply because the overlap between an $4f$ isotropic wave function and a 2D wave function decreases faster as the probe moves away from the Fe-As plane when compared to a more 3D wave function with $z$ components.

Moreover, we can now roughly simulate each Fe t$_{2g}$ orbital contributions to the FS using this phenomenological model. Starting with an isotropic Fe 3$d$ configuration, i.e., a normalized population of $1/\sqrt{3}$ for each considered 3$d$ orbital, we are able to tune the $xy$ orbital population - represented by the $B_{xy}$ coefficient - from $1/\sqrt{3}$ up to 1.
The inset of Fig. 5a shows $\langle \Psi_{4f}^{Eu}|\Psi_{3d}^{Fe}\rangle^{2}$ at a particular Eu-Fe distance $z_{0}=4.0$, chosen to better illustrate the behaviour mentioned above.  It is clear that, by increasing the $xy$ orbital contribution, the $\langle \Psi_{4f}^{Eu}|\Psi_{3d}^{Fe}\rangle^{2}$ diminishes drastically, in agreement with the behaviour presented by the Korringa rate $b$ in Fig. 4a for the transition metal $M$ substitution into Ba$_{0.8}$Eu$_{0.2}$Fe$_{2-y}M_{y}$As$_{2}$. 

We now turn our attention to the particular case which the considered local probe is in the Fe-As plane  ($M$ = Cu). In this case, we are able to calculate $\langle\Psi_{3d}^{Cu}|\Psi_{3d}^{Fe}\rangle^{2}$ as a function of the radius $r$ from a Fe atom. Fig. 5b displays the calculated squared overlap between the Cu 3$d$ and the Fe 3$d$ wavefunctions with the same t$_{2g}$ orbital configurations of Fig. 5a. However, when the planar $xy$ orbital component is the dominant contribution (solid red line) in this case, the overlap $\langle\Psi_{3d}^{Cu}|\Psi_{3d}^{Fe}\rangle^{2}$ decreases slower than that for the $xz/yz$ configuration. The inset of Fig. 5b shows an exponential increase of this squared overlap at $r = 3.0$ atomic units as a function
of the $xy$ weigh, indicating that higher planar contributions leads to higher Korringa rate $b$, in good agreement with our
experimental data shown in Fig. 4b.

In summary, we have employed a sensitive microscopic spin probe to study the suppression of the SDW magnetic phase
in BaFe$_{2}$As$_{2}$ via transition metal substitution. We demonstrated that the 3$d$ electrons tend to localize at the Fe-As plane as the SDW phase is suppressed due to changes in the structural parameters, independent on the transition metal substitution of choice. When the Fe-As distances are lowered either by chemical substitution or hydrostatic pressure, the crystal field splittings change and, in turn, the planar 
($xy$/$x^{2}-y^{2}$) orbital character of the Fe 3$d$ bands are strongly enhanced at the Fermi level. These effects and consequent orbital differentiation corroborate to the suppression of the itinerant SDW magnetic order and the subsequent emergence of the magnetic-mediated superconductivity in these materials.

This work was supported by FAPESP, AFOSR MURI, CNPq, FINEP-Brazil and NSF-USA.

\subsection{Methods}

Single crystals of Ba$_{1-x}$Eu$_{x}$Fe$_{2-y}M_{y}$As$_{2}$ ($M=$ Mn, Co, Cu, Ru,
and Ni) were grown using In-flux as described in Ref. \cite{Garitezi}.
The single crystals were checked by X-ray powder diffraction and submitted
to elemental analysis using a commercial Energy Dispersive Spectroscopy
(EDS) microprobe.  
No In-incorporation was detected in the crystals. From EDS analysis, we have extracted the actual $x_{Eu}$ and $y_M$ concentrations used throughout the text. The in-plane resistivity was measured using a four-probe method. Specific heat data were taken in a commercial small-mass calorimeter and the magnetization data was collected using a superconducting quantum
interference device (SQUID) magnetometer.
X-Band ($\nu=9.34$ GHz) ESR measurements were carried out in a commercial Bruker spectrometer with a continuous He gas-flow cryostat.

%
%
%
%
%


\begin{thebibliography}{9}
\bibitem{note} Here `$bottleneck$", ``$dynamic$" effects, and multiple bands effects are
not present \cite{ref20}. Those are reasonable assumptions because when ``$dynamic$" effects are present the $g$-values are usually strongly T-dependent,
when the bottleneck effect is relevant the Korringa rate $b$ decreases with the increasing of the concentration of the magnetic ions \cite{ref20}, and
when multiple band effects are present the calculated g-shift is much smaller than the measured one. None of these 
effects were observable in previous works \cite{Rosa} and neither in our ESR data of Fig.3 in the studied T-range.
\bibitem{shift} As defined in Eq. 2, $\Delta g \equiv g - g_{insulator}$. For Eu$^{2+}$,
$g_{insulator}$ is well known to be $1.993(1)$ \cite{Abragam}. For Mn$^{2+}$, an average of $g$ value in similar insulating compounds leads to $2.00(5)$ \cite{Abragam}.  For Cu$^{2+}$, we assume that the magnetic moment is entirely due to the spin and the $g_{insulator} \sim 2$. Although this assumption	leads	to a significant error in $J_{fs}(0)$, it does not affect the Korringa-like analysis (Eq. 1).
\bibitem{model} More specifically, the Fe 3d wave-function is written as
$\Psi_{3d} = R_{3d}(x,y,z) [B_{XZYZ} (\Phi_{XZ}(x,y,z) + \Phi_{YZ}(x,y,z))+ B_{XY}\Phi_{XY}]$,	where
$R_{3d}(x,y,z)$	is the $3d$ radial wave- function and $\Phi(x,y,z)$ are the cubic harmonics for each orbital.
\end{thebibliography}
\end{document}